\documentclass[preprint,aps,showpacs,superscriptaddress,tightenlines,a4paper]{revtex4}

\usepackage{amssymb}
\usepackage{amsmath}
\usepackage{bbold}
\usepackage{mathrsfs}
\usepackage{paralist}
\usepackage{graphicx}
\usepackage{xcolor}

\begin{document}
%
%: TITLE
%
    \title{The KdV hierarchy in optics}
%
%: AUTHORS
%
	\author{S. A. R. Horsley}
	\affiliation{Department of Physics and Astronomy, University of Exeter,
Stocker Road, Exeter, EX4 4QL}
	\email{s.horsley@exeter.ac.uk}
%
%: ABSTRACT
%
    \begin{abstract}
    There is a well explored relationship between quantum mechanical scattering from a potential and the Korteweg--de Vries (KdV) equation of fluid dynamics: if the potential is `evolved' according to the KdV equation then it will have the same reflectivity and transmissivity as a function of energy, for each snapshot in time.  In this work we explore this connection in optics, where the permittivity plays the role of the potential.  We begin by deriving the relationship between the Helmholtz equation and the KdV equation in terms of the current induced in a material when a permittivity profile is changed slightly.  It is then shown that the KdV equation can be used to design a plethora of bounded complex potentials that are relfectionless from both sides for all angles of incidence, and planar periodic media that exhibit a real Bloch vector for all angles of propagation.  Finally we apply the KdV equation to reduce the reflection of a wave from an interface between two media of differing refractive indices.
    \end{abstract}
%
% 03.50.De - Classical electromagnetism, Maxwell's equations.
% 81.05.Xj, 78.67.Pt - Metamaterials
%
    \pacs{03.50.De,81.05.Xj, 78.67.Pt}
    \maketitle
%
%: START CONTENT
%
%
%: Introduction
%
\section{Introduction}
\par
Wave propagation through inhomogeneous materials is more subtle than an application of ray optics would suggest.  When the material properties change on a scale that is comparable to the wavelength then the wave will reflect, and in general the reflection depends in an intricate way on the exact spatial dependence of the material properties.  Until relatively recently it was difficult to explore these subtle interactions of waves with matter, simply because the material properties could not be precisely specified as a function of position.  However, this situation has now changed somewhat; `metamaterial' structures~\cite{smith2006,cui2009,jahani2016} have been developed, where the material is engineered on a sub--wavelength scale so that it can be treated as a continuous function of position.  Such structures have been developed for controlling electromagnetic~\cite{schurig2006}, acoustic~\cite{cummer2016}, and water waves~\cite{hu2011}, as well as diffusion phenomena such as heat~\cite{han2014}.  To determine the necessary material properties to manipulate the wave in a desired way, one applies theories such as transformation optics~\cite{pendry2006,leonhardt2010}, which exploit the equivalence between inhomogeneous material properties and coordinate transformations.
\par
The purpose of this paper is to explore how a different equivalence might be used to understand the effect of inhomogeneous media on the propagation of waves.  We shall illustrate how a family of non--linear wave equations (known as the Korteweg--de Vries (KdV) hierarchy~\cite{drazin1989}) can be used to manipulate the reflectivity of a planar electromagnetic material through `evolving' the permittivity profile \(\epsilon(x)\) from one functional form to another.  Although the literature on the relationship between the KdV hierarchy and the Helmholtz equation is vast (see e.g.~\cite{bona2002} and references therein), the typical concern is with using the Helmholtz equation as a tool for solving the KdV equation rather the reverse, and the possibility of investigating this relationship with metamaterial structures does not seem to have been considered.  Here we imagine a regime that seems to be possible with metamaterials, where we have a very fine control over an isotropic permittivity \(\epsilon(x)\) as a function of position, in comparison to the wavelength of interest, and that we can control both its real and imaginary parts.
\par
In the first part of this work we re--derive the result that `evolving' a permittivity profile \(\epsilon(x)\) in `time' according to the KdV equation leads to a continuous family of profiles that all have a reflectivity that is different only by a phase.  This derivation is carried out in terms of the current that must be induced in a material when the permittivity profile is changed slightly.  Requiring that the radiation at \(x\) produced by this current only change the field in a way that depends on field values infinitesimally close to \(x\) leads us to the first two evolution equations in the KdV hierarchy.  We then show that some recent results concerning reflectionless media~\cite{horsley2015,longhi2015,longhi2016,horsley2016} can be derived in the same way.
\par
The second part of the paper is concerned with the possible application of the KdV hierarchy to the design of planar permittivity profiles.  We demonstrate the design of a family of complex materials that are reflectionless from both sides for all angles of incidence, and periodic planar media that do not exhibit a band gap for any angle of propagation.  Finally we demonstrate that the KdV equation can be used to modify an interface between two different values of the permittivity in such a way that the reflectivity is reduced for a range of angles of incidence.  Throughout this work we treat waves of a single frequency, and interest ourselves in manipulating the reflection of a planar medium as a function of angle.
%
%: Section 1
%
\section{Trajectories through equivalent inhomogeneous media\label{trajectory-section}}
\par
\begin{figure}[h!]
	\includegraphics[width=8cm]{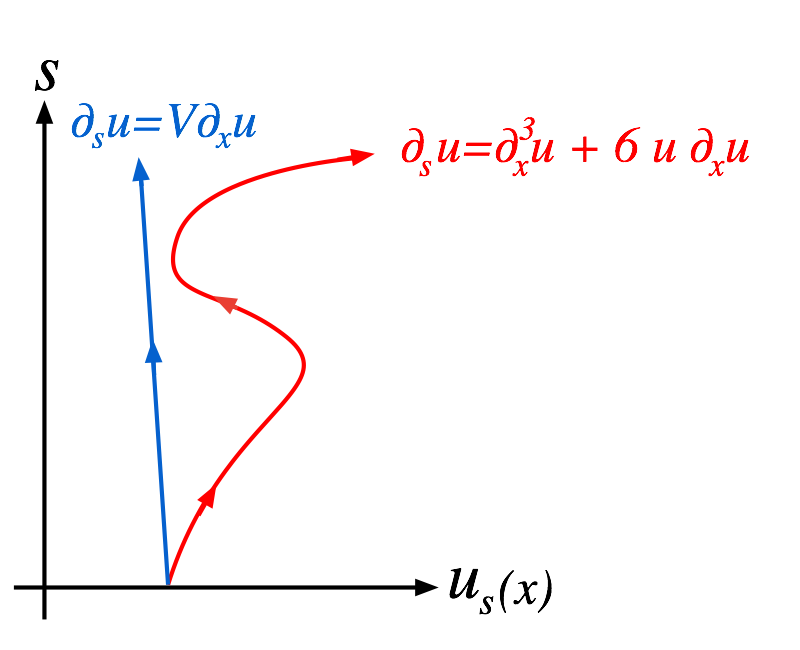}
	\caption{A schematic illustration of `permittivity space'.  The possible permittivity profiles \(\epsilon_{s}=1+u_{s}\) are parameterised by a `time' \(s\) and we move through `permittivity space' so that the field changes according to (\ref{local-operator}): \(\partial_{s}\varphi_{s}(x)=\hat{A}_{s}(x)\varphi_{s}(s)\).  This results in a family of generally complex permittivity profiles with identical transmission coefficients, and reflection coefficients that change according to (\ref{rev}).  An example of such an evolution is given by the Korteweg--de Vries equation (\ref{kdvequation}).\label{schematic_fig}}
\end{figure}

Given the lack of many general statements one can make about wave propagation through inhomogeneous media, one line of attack is to separate out the possible functions \(\epsilon(x)\) into families that have closely related scattering properties.  Take for example monochromatic electromagnetic waves polarized along \(\hat{\boldsymbol{z}}\) and propagating in the \(x\)--\(y\) plane through an inhomogeneous slab with complex permittivity \(\epsilon_{s}(x)=1+u_{s}(x)\).  These are governed by the Helmholtz equation
\begin{equation}
	\left[\frac{\partial^{2}}{\partial x^{2}}+k_{0}^{2}u_{s}(x)+k_{0}^{2}-k_{y}^{2}\right]\varphi_{s}(x)=0\label{helmholtz}
\end{equation}
where \(\varphi_{s}\) is the electric field amplitude, \(k_{y}=k_{0}\sin(\theta)\) is the in--plane wavevector determining the angle of incidence \(\theta\), and \(u_{s}(x)\to0\) as \(|x|\to\infty\).  \emph{The subscript `\(s\)' on the permittivity labels one of a continuous family of inhomogeneous media} (schematic shown in figure~\ref{schematic_fig}).  Differentiating (\ref{helmholtz}) with respect to \(s\) we can determine how the field changes as we move along a trajectory through this family of profiles
\begin{equation}
	\frac{\partial\varphi_{s}(x)}{\partial s}=h_{s}(x)-k_{0}^{2}\int dx'G_{s}(x,x',k_{0})\frac{\partial u_{s}(x')}{\partial s}\varphi_{s}(x')\label{field-change}
\end{equation}
where \(h_{s}(x)\) is some combination of the two solutions to the homogeneous equation (\ref{helmholtz}) that we are free to choose and \(G_{s}(x,x',k_{0})\) is the retarded Green function, obeying the equation
\begin{equation}
	\left[\frac{\partial^{2}}{\partial x^{2}}+k_{0}^{2}u_{s}(x)+k_{0}^{2}-k_{y}^{2}\right]G_{s}(x,x',k_{0})=\delta(x-x')\label{green}
\end{equation}
%
%: Schematic figure
%
\noindent
Equation (\ref{field-change}) tells us that the field at a fixed position \(x\) in the profile \(u_{s+ds}(x)\) differs from that in \(u_{s}(x)\) generally depending on all other points in space.  While this is not at all informative in the general case, this paper is concerned with those particular choices of evolution equation \(\partial u_{s}/\partial s\) when the right hand side of (\ref{field-change}) can be made to depend only on \(x\)
\begin{equation}
	\frac{\partial\varphi_{s}(x)}{\partial s}=\hat{A}_{s}(x)\varphi_{s}(x)\label{local-operator}
\end{equation}
with \(\hat{A}_{s}(x)\) some operator depending on \(\epsilon_{s}(x)\) and derivatives with respect to \(x\) (it is a local operator).  To better understand the physical meaning of reducing (\ref{field-change}) to (\ref{local-operator}), suppose we understand \(j=(\partial_{s}u_{s})\varphi_{s}\) as a current, which is the source of radiation \(\partial_{s}\varphi_{s}\) in (\ref{field-change}).  Our choice of \(\partial_{s}u_{s}\) that leads to (\ref{local-operator}) amounts to a choice of envelope function within the current distribution \(j\) such that the radiation that reaches \(x\) from all other points in space cancels out, leaving only that coming from infinitesimally close to \(x\).  That this is at all possible in any non--trivial cases is surprising, and as we shall see sometimes quite useful.
\par
Equation (\ref{local-operator}) is of the same form as the time dependent Schr\"odinger equation, and by analogy the general solution is a path ordered exponential~\cite{renteln2014}.   An important aspect of the theory we are discussing is that in many cases \(\hat{A}_{s}\) can be reduced to something independent of `time' (`\(s\)') when \(|x|\to\infty\), because here the permittivity reduces to unity.  The path ordering then ceases to matter and we have \(\varphi_{s}(x)=\exp(\hat{A}s)\varphi_{0}(x)\).  Now consider a wave incident from the left of the profile.  On the far left we have an incident wave plus a reflected one, \(\varphi_{0}=\exp({\rm i}k_{x}x)+r_{0}\exp(-{\rm i}k_{x}x)\).  The `evolution' over \(s\) changes this to \(\varphi_{s}=\exp(A({\rm i}k_{x})s)\exp({\rm i}k_{x}x)+r_{0}\exp(A(-{\rm i}k_{x})s)\exp(-{\rm i}k_{x}x)\), where \(k_{x}=[k_{0}^{2}-k_{y}^{2}]^{1/2}\) and \(A({\rm i}k_{x})\) is the function obtained from replacing the derivatives within the operator by \({\rm i}k_{x}\).  Therefore in cases where (\ref{local-operator}) holds for all \(x\), the reflection coefficient from the profile \(r_{s}\) `evolves' in `time' as follows
\begin{equation}
	r_{s}={\rm e}^{-\left[A({\rm i}k_{x})-A(-{\rm i}k_{x})\right]s}r_{0}\label{rev}
\end{equation}
and the transmission coefficient is left unchanged
\begin{equation}
	t_{s}=t_{0}.\label{tev}
\end{equation}
This paper is concerned with the application of (\ref{rev}--\ref{tev}) to manipulate the reflection from inhomogeneous complex media.
%
%: Translational symmetry
%
\subsection{Translational symmetry\label{transsec}}
\par
We haven't yet shown that the change in the field (\ref{field-change}) can ever be written as the local operation (\ref{local-operator}).  This can be demonstrated in a rather elementary case, where the trajectory parameterized by `\(s\)' simply corresponds to a translation of the permittivity profile in space.
\par
A family of permittivity profiles that are all of the same shape, but centred at different positions can be generated by
\[
	\frac{\partial u_{s}(x)}{\partial s}=V\frac{\partial u_{s}(x)}{\partial x}
\]
where \(V\) is the `velocity' at which the profile moves as a function of \(s\).  Substituting this into (\ref{field-change}), integrating by parts, and applying (\ref{helmholtz}) and (\ref{green}) we obtain
\begin{multline}
	\frac{\partial\varphi_{s}(x)}{\partial s}=V\frac{\partial\varphi_{s}(x)}{\partial x}+h_{s}(x)\\
	-V\int dx'\frac{\partial}{\partial x'}\left[\frac{\partial G_{s}(x,x',k_{0})}{\partial x'}\frac{\partial\varphi_{s}(x')}{\partial x'}+(k_{0}^{2}-k_{y}^{2})\varphi_{s}(x')G_{s}(x,x',k_{0})\right].
\end{multline}
The boundary terms from the integral correspond to waves whose source lies at infinity; solutions to the homogeneous equation (\ref{helmholtz}).  Therefore \(h_{s}\) can always be chosen to eliminate these terms and we are left with
\begin{equation}
	\frac{\partial\varphi_{s}(x)}{\partial s}=V\frac{\partial\varphi_{s}(x)}{\partial x}\label{field-translation}
\end{equation}
the right hand side of which is of the form (\ref{local-operator}), with \(\hat{A}=V\partial_{x}\).  Thus the reflection coefficients change with \(s\) according to
\begin{equation}
	r_{s}={\rm e}^{-2{\rm i}k_{x}V s}r_{0}\label{rtrans}
\end{equation}
with the transmission coefficients unaltered.  This is the expected change in the reflection coefficient after a translation of the permittivity by a distance \(Vs\).  Note that \(V\) is not restricted to real values, and that a translation of the profile by a complex distance will exponentially amplify or diminish the reflection rather than shift it by a phase~\cite{horsley2016}.
%
%: KdV equation
%
\subsection{The Korteweg--de Vries hierarchy\label{kdvsec}}
%
%: KdV Figure
%
\begin{figure}[h!]
	\includegraphics[width=16cm]{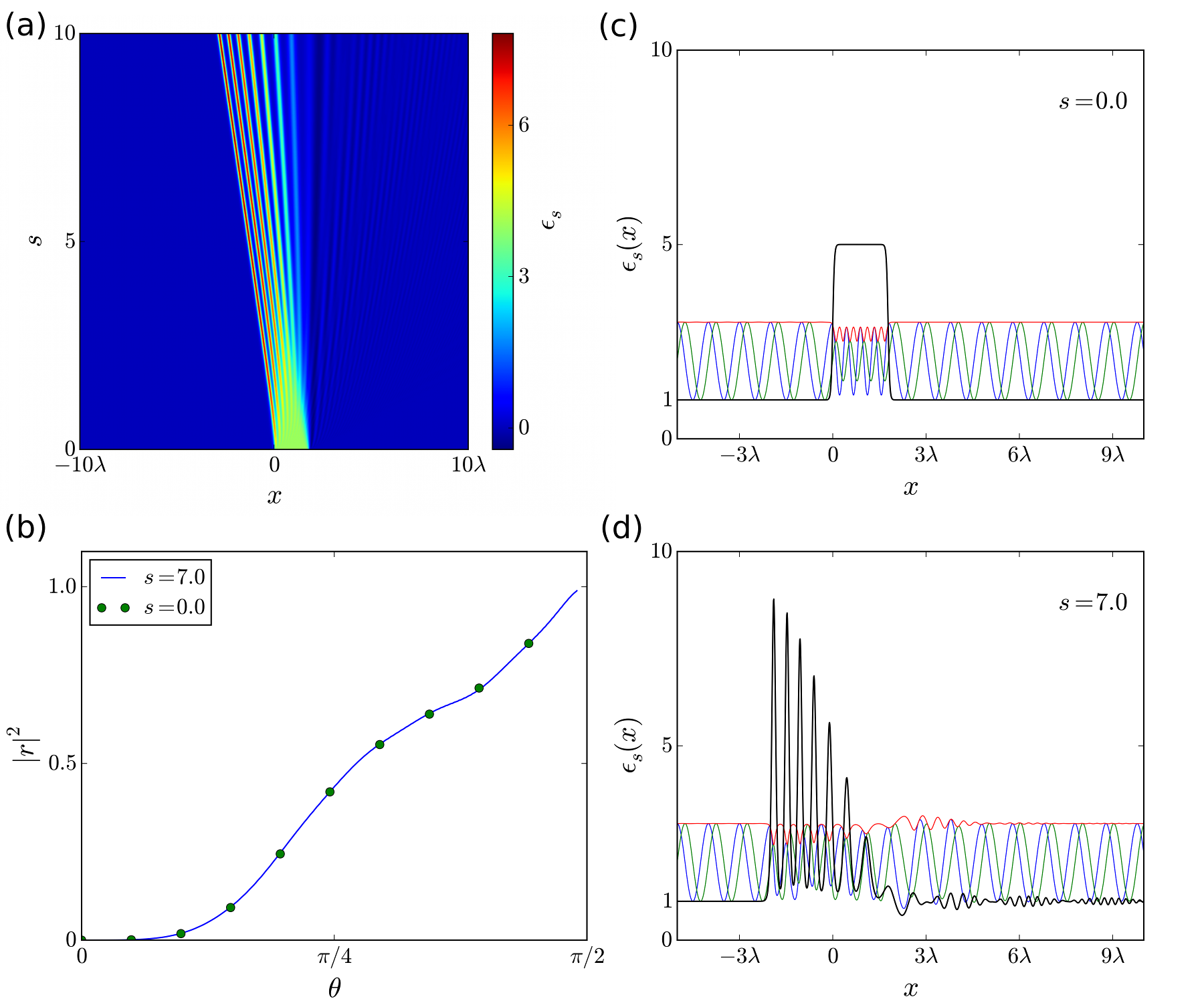}
	\caption{(a) Example profile evolving according to (\ref{kdvequation}) from an initial homogeneous slab \(u_{s=0}=[1+\tanh(3 k_0 x)][1+\tanh(3 k_0(3.975\lambda/\sqrt{5}-x))]\), for \(s\in[0,10]\), with \(a=1/8k_0^3\) and \(\lambda=2\pi/k_{0}\) (numerical integration of the KdV equation was adapted from the \emph{Scipy Cookbook}~\cite{scipy2001}).  As \(s\) increases the slab breaks up into small ripples that rapidly spread out, plus solitons~\cite{grunert2009}.  This particular profile is chosen to have negligible reflection at normal incidence, and panels (c--d) show that this property is retained as \(s\) increases.  The real and imaginary parts of the field are superimposed on the permittivity profile in blue and green respectively, and the absolute value is shown in red.  (b) Reflectivity as a function of angle \(k_{y}=k_{0}\sin(\theta)\) for the two permittivity profiles shown in (c--d).\label{kdvfig}}
\end{figure}

\par
There are an infinite number of less elementary evolution equations for \(u_{s}\) that reduce the change in the field (\ref{field-change}) to a local operation (\ref{local-operator}).  For example suppose the evolution of the permittivity is governed by some third order differential equation \(\partial_{s}u_{s}=a\partial_{x}^{3}u_{s}+\dots\).  With this form of evolution~\footnote{A choice of even order differential equation, e.g. \(\partial_{s} u_{s}=a\partial_{x}^{2} u_{s}+\dots\) does not yield a non--trivial equation of the form (\ref{local-operator}).}, the change in the field (\ref{field-change}) reduces to
\begin{multline}
	\frac{\partial\varphi_{s}}{\partial s}=h_{s}(x)+ak_{0}^{2}\left[2u_{s}(x)\frac{\partial\varphi_{s}(x)}{\partial x}-\frac{\partial u_{s}(x)}{\partial x}\varphi_{s}(x)\right]
	\\[10pt]+2ak_{0}^{2}\int dx'\left[3k_{0}^{2}u_{s}(x')\frac{\partial u_{s}(x')}{\partial x'}+2(k_{0}^{2} - k_{y}^{2})\frac{\partial u_{s}(x')}{\partial x'}\right]\varphi_{s}(x')G_{s}(x,x',k_{0})+\dots\label{d3x}
\end{multline}
the integral term in (\ref{d3x}) can be removed through---as in the preceding section---choosing the homogeneous part of the field \(h_{s}\) to eliminate any boundary terms, and choosing the evolution equation for \(u_{s}\) to be,
\begin{equation}
	\frac{\partial u_{s}(x)}{\partial s}=a\left[\frac{\partial^{3} u_{s}(x)}{\partial x^{3}}+6k_{0}^{2}u_{s}(x)\frac{\partial u_{s}(x)}{\partial x}\right]\label{kdvequation}
\end{equation}
which is the Korteweg--de Vries (KdV) equation~\cite{korteweg1895}, a non--linear wave equation that was initially applied to the propagation of shallow water waves, but has since appeared in many areas of physics, including plasma physics~\cite{shukla2001} and acoustics~\cite{naugolnykh1998}. If the permittivity profile is subject to evolution by (\ref{kdvequation})---which is notably independent of \(k_{y}\)---the field correspondingly transforms as
\begin{equation}
	\frac{\partial\varphi_{s}(x)}{\partial s}=4a\frac{\partial^{3}\varphi_{s}(x)}{\partial x^{3}}+6ak_{0}^{2}u_{s}(x)\frac{\partial\varphi_{s}(x)}{\partial x}+3k_{0}^{2}a\frac{\partial u_{s}(x)}{\partial x}\varphi_{s}(x)
\end{equation}
which has the form (\ref{local-operator}) with \(\hat{A}_{s}=4a\partial_{x}^{3}+6ak_{0}^{2}u_{s}\partial_{x}+3ak_{0}^{2}\partial_{x}u_{s}\).  In the limit \(|x|\to\infty\) this reduces to \(\hat{A}=4a\partial_{x}^{3}\), meaning that as we increase `\(s\)' the reflection coefficients change as follows (see (\ref{rev}))
\begin{equation}
	r_{s}=\exp(8{\rm i}ak_{x}^{3}s)r_{0}\label{rkdv}
\end{equation}
with the transmission coefficients unchanged.  Figure~\ref{kdvfig} demonstrates the validity of (\ref{rkdv}) for real \(a\), showing that a slab of uniform real permittivity evolved in `time' according to the KdV equation becomes a complicated sum of solitary waves, \emph{while retaining the same reflectivity as a function of angle}.  It is interesting to note that a wave evolving according to the KdV equation satisfies an infinite number of conservation laws~\cite{drazin1989}, the first two of which are \(\partial_{s}\int u_{s} dx=0\) and \(\partial_{s}\int u_{s}^{2}dx=0\): for us this translates into an infinite number of integrals of functions of the permittivity that must remain identical for the medium to have the same reflectivity as a function of angle. 
\par
\begin{figure}[h!]
	\includegraphics[width=16cm]{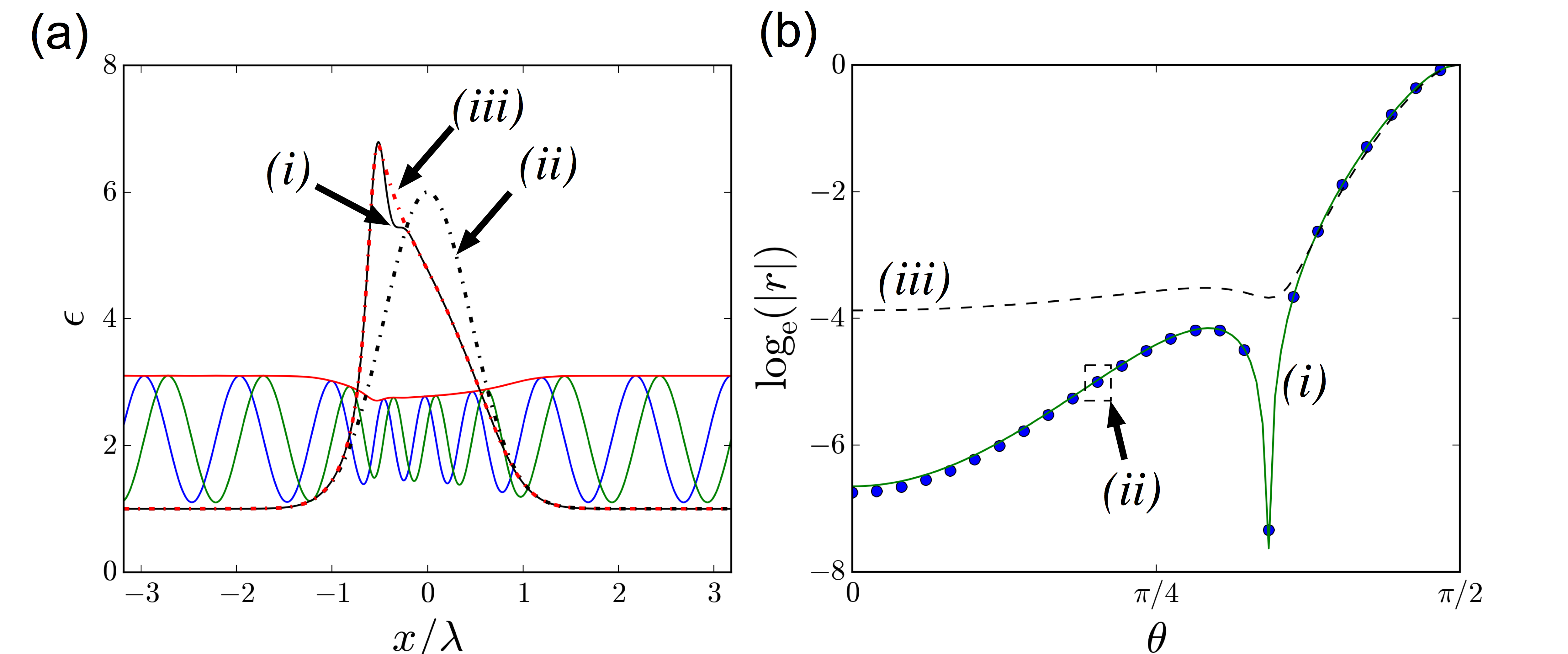}
	\caption{Increasing the slope of a permittivity profile usually increases its reflectivity.  The fifth order KdV equation (\ref{kdv5}) (for example) can generate profiles where the leading edge is steepened while retaining the same low reflectivity as a function of angle. (a) An initial gaussian profile (ii) (dashed black) is evolved into (i) (solid black) according to the fifth order KdV equation.  Wave propagation is then compared within a very similar profile (iii) (dashed red).  Superimposed is a plot of wave propagation through profile (i) at normal incidence (colours as in figure~\ref{kdvfig}). (b) The log of the modulus of the reflection coefficient as a function of angle plotted for the three profiles (i--iii).  The steepened profile (i) retains the same low reflectivity as the initial gaussian (ii), while a very similar profile (iii) has a consistently higher reflectivity.\label{kdv5fig}}
\end{figure}

A similar procedure can be carried out to generate a bewildering variety of evolution equations for \(u_{s}\), some of which are unavoidably dependent on \(k_{y}\) and some not.  Besides patience, what one learns is; firstly that there are a vast number of permittivity profiles with identical reflectivity as a function of angle; and less surprisingly that there is an even more vast set of profiles with the same reflectivity at a fixed angle of incidence, all of which can be generated in a systematic way.  As a further example we take \(\partial_{s}u_{s}=a\partial_{x}^{5}u_{s}+\dots\) in (\ref{field-change}) and apply (\ref{helmholtz}) and (\ref{green}), obtaining after a long series of manipulations the following \(k_{y}\) independent evolution equation for the permittivity
\begin{equation}
	\frac{\partial u_{s}(x)}{\partial s}=a\left[\frac{\partial^{5}u_{s}(x)}{\partial x^{5}}+10k_{0}^{2}u_{s}(x)\frac{\partial^{3}u_{s}(x)}{\partial x^{3}}+20k_{0}^{2}\frac{\partial u_{s}(x)}{\partial x}\frac{\partial^{2} u_{s}(x)}{\partial x^{2}}+30k_{0}^{4}u_{s}^{2}(x)\frac{\partial u_{s}(x)}{\partial x}\right]\label{kdv5}
\end{equation}
with the field evolving according to
\begin{multline}
	\frac{\partial\varphi_{s}(x)}{\partial s}=16a\frac{\partial^{5}\varphi_{s}(x)}{\partial x^{5}}+40ak_{0}^{2}u_{s}(x)\frac{\partial^{3}\varphi_{s}(x)}{\partial x^{3}}+60ak_{0}^{2}\frac{\partial u_{s}}{\partial x}\frac{\partial^{2}\varphi_{s}(x)}{\partial x^{2}}\\[10pt]
	+10ak_{0}^{2}\left[5\frac{\partial^{2}u_{s}(x)}{\partial x^{2}}+3k_{0}^{2}u_{s}^{2}(x)\right]\frac{\partial\varphi_{s}(x)}{\partial x}+15ak_{0}^{2}\left[\frac{\partial^{3}u_{s}}{\partial x^{3}}+2k_{0}^{2}u_{s}(x)\frac{\partial u_{s}(x)}{\partial x}\right]\varphi_{s}(x)
\end{multline}
In this case in the limit \(|x|\to\infty\) the local operator reduces to \(\hat{A}=16a\partial^{5}_{x}\), meaning that the reflection coefficients transform as \(r_{s}=\exp(-32{\rm i}ak_{x}^{5}s)r_{0}\) (see figure~\ref{kdv5fig}).  The above two evolution equations (\ref{kdvequation}) and (\ref{kdv5}) are simply the first two of what is known as the `\emph{KdV hierarchy}'~\cite{conte1999}, an infinite set of increasingly complicated non--linear equations under the evolution of which the reflection coefficient transforms as (\ref{rev}).
\subsection{Kramers--Kronig media}
\par
Recent work~\cite{horsley2015,longhi2015,longhi2016,horsley2016} has shown that complex inhomogenous media satisfying the spatial Kramers--Kronig relations have curious scattering properties, being generally reflectionless from one side and in many cases having unit transmission.  Although not initially cast as such, this is also a case when the `evolution' of the field can be written asymptotically as a local operation (\ref{local-operator}), which we now briefly describe.  
\par
Suppose that our initial permittivity profile is vacuum \(u_{s}=0\), and that we consider a right--going wave \(\varphi_{0}(x)=\exp({\rm i}k_{x}x)\).  In this case the change in the field (\ref{field-change}) is equal to
\begin{equation}
	\frac{\partial\varphi_{s}(x)}{\partial s}=-\frac{k_{0}^{2}}{2{\rm i}k_{x}}\left[{\rm e}^{{\rm i}k_{x}x}\int_{-\infty}^{x}\frac{\partial u_{s}(x_1)}{\partial s}dx_1+{\rm e}^{-{\rm i}k_{x}x}\int_{x}^{\infty}\frac{\partial u_{s}(x_1)}{\partial s}{\rm e}^{2{\rm i}k_{x}x_1}dx_1\right]\label{field-derivative-0}
\end{equation}
where we imposed \(h_{s}=0\) and used the free space Green function: \(G(x,x',k_{0})=\exp({\rm i}k_{x}|x-x'|)/2{\rm i}k_{x}\).  We now consider the position as a complex variable \(x=x'+{\rm i}x''\), and take \(\partial u_{s}/\partial s\) as an analytic function that tends to zero as \(|x|\to\infty\) in the upper half complex position plane.  As we take (\ref{field-derivative-0}) towards infinity in the upper half plane the first term in the square brackets decays exponentially to zero as \({\rm e}^{-k_{x}x''}\).  The second term decays with the same exponent, as is clear if we successively integrate by parts to obtain e.g. for \(N\) integrations
\begin{multline}
	\int_{x}^{\infty}\frac{\partial u_{s}(x_1)}{\partial s}{\rm e}^{2{\rm i}k_{x}x_1}dx_1={\rm e}^{2{\rm i}k_{x}x}\sum_{n=0}^{N}(-1)^{n+1}\left(\frac{1}{2{\rm i}k_{x}}\right)^{n+1}\frac{d^{n}}{dx^{n}}\left(\frac{\partial u_{s}(x)}{\partial s}\right)\\
	+(-1)^{N+1}\left(\frac{1}{2{\rm i}k_{x}}\right)^{N+1}\int_{x}^{\infty}\frac{d^{N}}{d x_{1}^{N}}\left(\frac{\partial u_{s}(x_1)}{\partial s}\right){\rm e}^{2{\rm i}k_{x}x_1}dx_1
\end{multline}
for large \(N\) the integral term on the second line becomes negligibly small as we move far into the upper half plane.  Therefore \(\partial\varphi_{s}/\partial s\) is analytic and asymptotically tends to zero in the upper half complex position plane.  Given that functions that are analytic in one half of the complex position have one--sided Fourier (\(k\)) spectra~\cite{volume8} this means that we can change the permittivity away from vacuum by adding in an inhomogeneous part that is analytic in one half of the complex position plane and we will not generate any counter propagating waves: \emph{there will be no reflection from the profile}.  Indeed, this is not just true for small changes of the permittivity away from vacuum, 
but there is never any reflection generated if we continue to change \(u_{s}\) in such a way that it remains analytic in one half of the complex position plane.  To see this, consider the formula for the second derivative of \(\varphi_{s}\) with respect to \(s\)
\begin{equation}
	\frac{\partial^{2}\varphi_{s}}{\partial s^{2}}=-\frac{k_{0}^{2}}{2{\rm i}k_{x}}\left[{\rm e}^{{\rm i}k_{x}x}\int_{-\infty}^{x}dx_{1}U_{s}(x_{1})+{\rm e}^{-{\rm i}k_{x}x}\int_{x}^{\infty}dx_{1}U_{s}(x_{1}){\rm e}^{2{\rm i}k_{x}x}\right]\label{d2s}
\end{equation}
where
\[
	U_{s}(x_{1})=2\frac{\partial u_{s}(x_{1})}{\partial s}\frac{\partial \varphi_{s}(x_{1})}{\partial s}{\rm e}^{-{\rm i}k_{x}x}+\frac{\partial^{2}u_{s}(x_{1})}{\partial s^{2}}
\]
From our analysis of \(\partial \varphi_{s}/\partial s\) we can see that, so long as \(\partial^{2} u_{s}/\partial s^{2}\) is analytic and tends to zero in the upper half plane then \(U_{s}\) will also have this property.  Thus, given that (\ref{d2s}) is of the same form as (\ref{field-derivative-0}) we can conclude that \(\partial^{2}\varphi_{s}/\partial s^{2}\) is also analytic and asymptotically tending to zero in the upper half plane.  This procedure can be used to iteratively show that all derivatives of \(\varphi_{s}\) have this property.  Therefore if we start with a right--going wave in vacuum and `evolve' \(u_{s}\) such that stays analytic and asymptotically zero in the upper half plane then the change in the permittivity will never generate any reflection. 
%
%: KdV5 and KK figure
%
\par
One consequence of this analyticity is that, as in the case of the KdV hierarchy, on the far left and far right of the profile the `evolution' equation (\ref{field-derivative-0}) reduces to a local operation (\ref{local-operator})
\[
	\frac{\partial\varphi_{s}(x)}{\partial s}\to\varphi_{s}(x)\times\begin{cases}
		0&x\to-\infty\\
		\frac{{\rm i}k_{0}}{2}\int_{-\infty}^{\infty}\frac{\partial u_{s}(x_1)}{\partial s}dx_1&x\to+\infty
	\end{cases}
\]
(in cases where \(\partial u_{s}/\partial s\) decays to zero as \(1/x\), the integrals in (\ref{field-derivative-0}) must be evaluated as a principal value, taking them to be non--zero over \([-L,L]\) then taking \(L\to\infty\)).  Notice that at \(+\infty\) the operator \(\hat{A}_{s}\) given in (\ref{local-operator}) depends on `\(s\)' so that, except in restricted cases, the evolution equation for the transmission coefficient (\ref{tev}) no longer applies and the transmission coefficient evolves as
\begin{equation}
	t_{s}=\exp\left(\frac{{\rm i}k_{0}}{2}\int^{s}_{0}ds'\int_{-\infty}^{\infty}\frac{\partial u_{s'}(x_{1})}{\partial s'}dx_{1}\right)=\exp\left(\frac{{\rm i}k_{0}}{2}\int_{-\infty}^{\infty}u_{s'}(x_{1})dx_{1}\right)
\end{equation}
Longhi has recently pointed out the importance of the `cancellation condition', \(\int_{-\infty}^{\infty}u_{s}(x_{1})dx_{1}=0\) for the definition of plane wave scattering states at infinity~\cite{longhi2016}, and here we reproduce the finding that all profiles that satisfy this condition are invisible from one side: \emph{they do not reflect for any angle of incidence and transmit without a phase shift}~\cite{longhi2016,horsley2016}.  Moreover if such a profile is evolved according to one of the equations in the KdV hierarchy (see~\ref{kdvsec}) then it remains analytic in one half plane and always satisfies the cancellation condition, by virtue of the conservation law \(\partial_{s}\int u_{s}(x) dx=0\).  The condition that these profiles are analytic in one half plane means that they are always complex, and therefore have balanced regions where the wave is absorbed and amplified (recent work on parity--time symmetric materials~\cite{bender2007} has seen experimental realisations of such media~\cite{ruter2010,feng2013}).
%
%: The complex KdV equation
%
\section{Controlling reflection from complex profiles}
\par
The simple transformation of the reflection coefficients given by (\ref{rev}) enables us to use the evolution equations of the preceding sections to design complex permittivity profiles with prescribed reflection coefficients.
\begin{figure}[h!]
	\includegraphics[width=16cm]{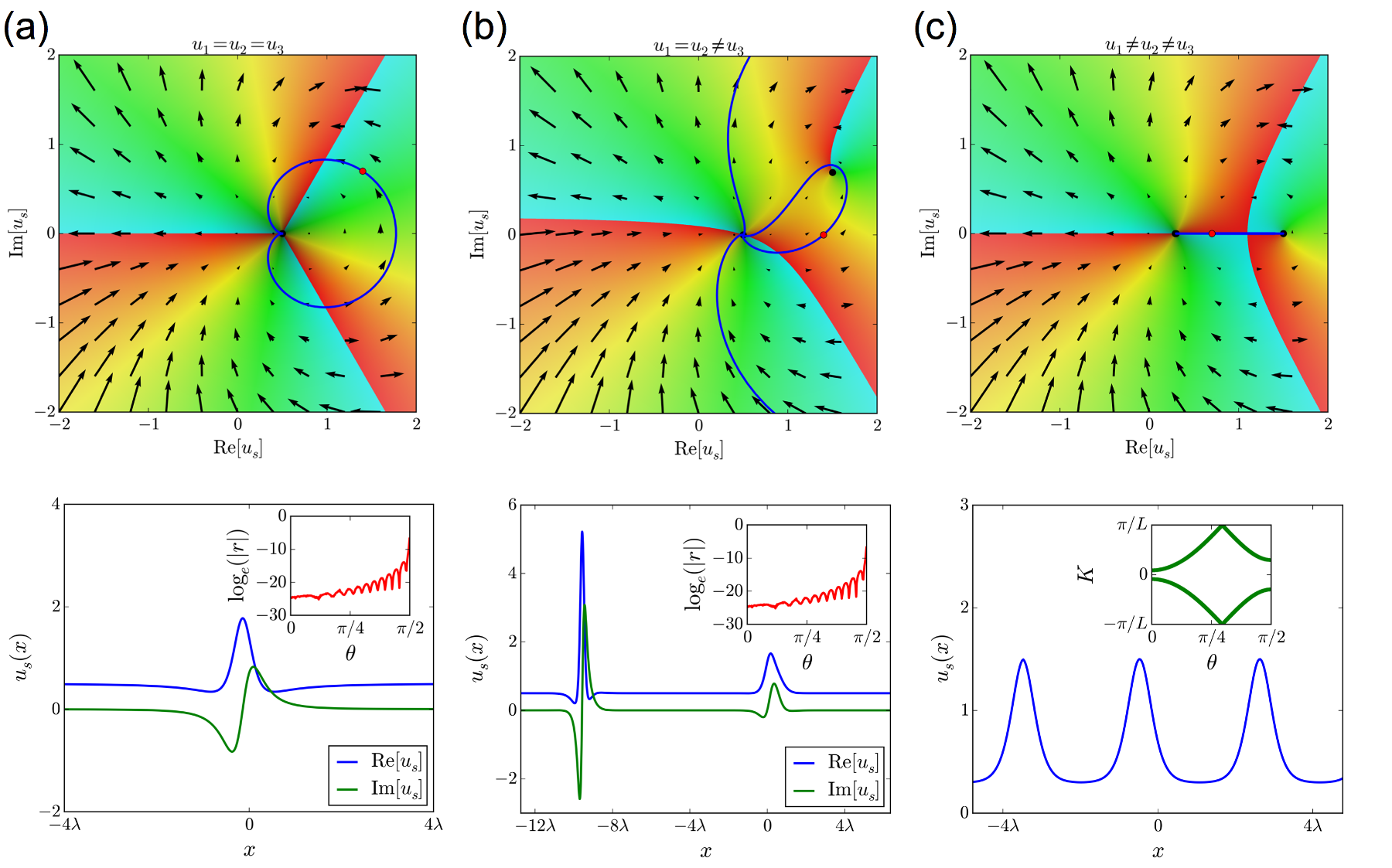}
	\caption{Examples of reflectionless permittivity profiles given by (\ref{complex_soliton}).  The figure shows instances of the three different classes of profile.  In each case the blue line in the upper panel shows the path traced out in \(u_{s}\) space by (\ref{complex_soliton}) with the background colour and arrows indicating the value of \(\partial u_{s}/\partial x\) and the red dot showing \(u_{s}(0)\) (note that the jumps in the background colour indicate branch cuts): (a) here the three roots of (\ref{complex_soliton}) coincide \(u_{1}=u_{2}=u_{3}=0.5\), which is the complex profile \(u_{s}(x)=u_{1}-2/[k_{0}^{2}(x+{\rm i}x_{0})^{2}]\), that exhibits a mixture of both loss and gain (see~\cite{longhi2015} and the appendix of~\cite{horsley2016} for a discussion of this particular profile and its practical limitations).  Inset in the lower plot is a logarithmic plot of the reflection coefficient as a function of angle;  (b) two of the three roots coincide \(u_{1}=u_{2}=0.5\) and \(u_{3}=1.5+0.7{\rm i}\) in this case resulting in a relatively complicated path in \(u\) space that in general gives rise to quite irregular---although invisible---profiles; (c) all three roots are distinct and real \(u_{1}=0.29\), \(u_{2}=0.3\), \(u_{3}=1.5\) resulting in a real valued periodic medium (period \(L=1.55\lambda\)) in the form of a \emph{cnoidal wave}~\cite{drazin1989}.  Inset in the lower plot is the Bloch wave--vector for a fixed frequency which is a real valued function of the propagation angle \(\theta\), illustrating the absence of any band gap (which is due to the lack of any reflection).\label{invisible-fig}}
\end{figure}
%
%: Invisible planar media
%
\subsection{Reflectionless planar media}
\par
In the case of real profiles it is well known that the solitons of the KdV hierarchy are reflectionless for all angles of incidence~\cite{drazin1989,volume3,lekner2010} (the reflectionless P\"oschl--Teller profile~\cite{poschl1933} is a soliton in the KdV hierarchy).  This is because a soliton moves at a constant velocity without changing shape.  Therefore it's evolution is simultaneously governed by both the symmetries of sections~\ref{transsec} and~\ref{kdvsec}.  But the reflection coefficient cannot evolve according to both (\ref{rtrans}) and (\ref{rkdv}) except if it is zero for all \(k_{y}\).  The same argument can also be carried out to derive complex permittivity profiles that have zero reflection coefficient from both sides, for all angles of incidence.  If we demand that the solution to the KdV equation (\ref{kdvequation}) translates in space at uniform velocity \(V\) over `time' \(s\) then the resulting profile will be reflectionless from both sides and have unit transmission, for all angles of incidence.  Substituting \(u_s=u_s(x+V t)\) we find that such a profile is a solution to
\[
	\frac{\partial^{3}u_{s}}{\partial x^{3}}+6k_{0}^{2}u_{s}\frac{\partial u_{s}}{\partial x}-\frac{V}{a}\frac{\partial u_{s}}{\partial x}=0
\]
which can be integrated three times to give
\begin{equation}
	x(u_{s})=\pm\int_{u_{s}(0)}^{u_{s}}\frac{d u}{{\rm i}k_{0}\sqrt{2(u-u_1)(u-u_2)(u-u_3)}}\label{complex_soliton}
\end{equation}
where the sign is chosen according to the branch of the square root and the cubic function in the denominator is given by
\begin{equation}
	(u-u_1)(u-u_2)(u-u_3)=u^{3}-\frac{V}{2ak_{0}^{2}}u^{2}-\kappa_{1}u-\kappa_{2}\label{roots}
\end{equation}
where \(\kappa_{1}\) and \(\kappa_{2}\) are integration constants.  In general the solution to the integral (\ref{complex_soliton}) can be written in terms of elliptic functions~\cite{drazin1989}, but we shall not use this representation here.
\par
Although there has been quite a lot done to understand the properties of the complex KdV equation (e.g.~\cite{birnir1987,bona2008}), the complex solutions of this equation do not seem to have been widely examined for their properties as inhomogeneous optical media.  In fact, with the freedom to choose \(u_{s}(0)\), the velocity, and the two integration constants, (\ref{complex_soliton}) defines a large number of reflectionless \emph{complex} permittivity profiles that include the soliton profiles as a special case.  The different profiles that are described by (\ref{complex_soliton}) can be divided up into three kinds based on the coincidence of the roots in (\ref{roots}); (i) \(u_{1}=u_{2}=u_{3}\), where the medium is complex and \(u_{s}(x)\) tends to the constant \(u_{1}\) at infinity; (ii) \(u_{1}=u_{2}\neq u_{3}\) where the spatial extent of the profile depends on the argument of \(u_{1}-u_{3}\), e.g. it is an infinite periodic medium when \({\rm arg}(u_{1}-u_{3})=0\), and confined to a finite region of space when \({\rm arg}(u_{1}-u_{3})=\pi\); and (iii) \(u_{1}\neq u_{2}\neq u_{3}\), which is a medium of infinite extent and may not be periodic.  Figure~\ref{invisible-fig} shows instances of each of these cases, demonstrating the ultra low reflectivity of two confined profiles and the lack of any band gap (as a function of angle) for a periodic medium.  A very peculiar feature of all of these profiles is that they retain their zero reflectivity and unit transmissivity as the angle of incidence is varied (which amounts to varying the wavelength inside the medium), but \emph{not} as the frequency is varied.  This is exemplified in the final example of figure~\ref{invisible-fig} where---at a fixed frequency---the `cnoidal Bragg mirror' exhibits no band gap as a function of angle, while if the frequency is varied from the chosen value a band gap does in general appear. 
\begin{figure}[h!]
	\includegraphics[width=14cm]{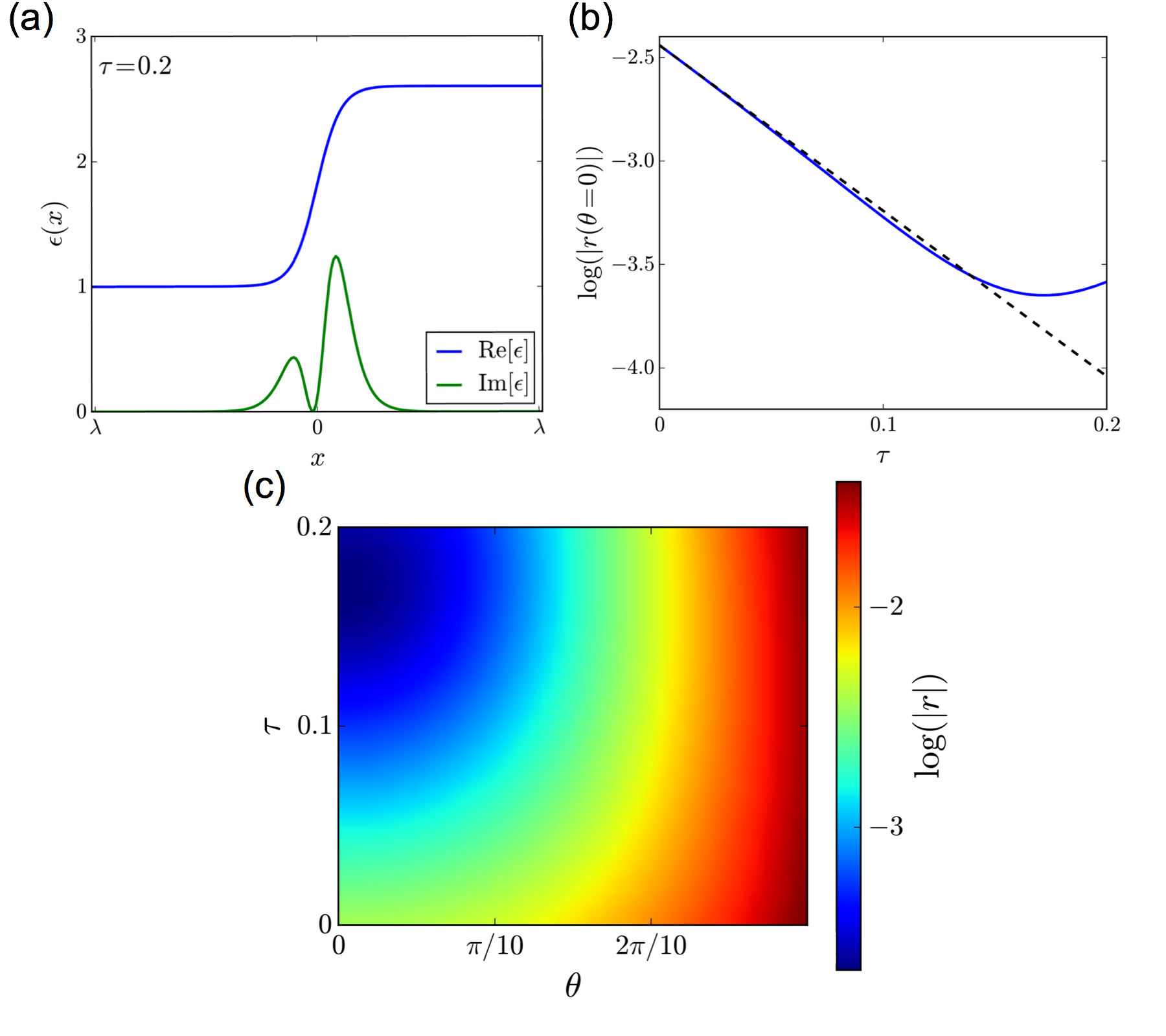}
	\caption{If the KdV equation is applied to evolve an initially permittivity profile for an imaginary `time' interval \(s={\rm i}\tau\) then the reflection from the profile will exponentially diminish or increase, depending on the sign of \(\tau\): \(r_{s}=\exp{(-8 a k_{x}^{3}\tau)}r_{0}\).  For small \(\tau\) the permittivity simply acquires an additional imaginary part; (a)  A smoothed out step profile acquires the imaginary part given in (\ref{evolved-profile}), here shown for \(\tau=0.2\); (b) The natural logarithm of the reflection coefficient plotted for increasing \(\tau\).  Superimposed is the dashed line showing the predicted decrease of reflectivity \(\log(r_{s})=\log(r_{0})-8 a k_{x}^{3}\tau\), which in this case is only expected to hold for small \(\tau\); (c) The logarithm of the reflectivity plotted as a function of both the angle of incidence and \(\tau\).  As is evident from (\ref{rkdv}), the exponential decrease in the reflectivity falls off as \(\cos^{3}(\theta)\) as we move away from normal incidence.\label{reflection-fig}}
\end{figure}
%
%: Reducing the reflection at an interface
%
\subsection{Reducing the reflection from an interface}
\par
The evolution equations discussed above can also be used to diminish reflection from a profile.  Suppose we have some given inhomogeneous medium and want to reduce its reflection for all angles.  To be concrete let us suppose a smoothed out step function
\begin{equation}
	u_{0}(x)=\frac{U_0}{2}\left[1+\tanh(x/\sigma)\right]\label{initial-profile}
\end{equation}
which represents a material \(\epsilon(x)=1+u_0(x)\) that is vacuum at \(x\to-\infty\) and has constant permittivity \(1+U_{0}\) at \(x\to+\infty\), with the transition between the two occurring over a length scale \(\sigma\) around \(x=0\).  It is often practically useful to reduce the reflection of a wave for a range of angles as it passes from one medium to another, while keeping the length scale of the transition constant.  Here we use the above theory to suggest some ways of doing this.
\par
As we have already shown, the reflection coefficient changes by an exponential factor under translation (\ref{rtrans}), or evolution by the KdV hierarchy (\ref{rkdv}).  For real values of the parameters this factor is simply a phase shift, leaving the reflectivity \(|r|^{2}\) unaffected.  Meanwhile for complex values of e.g. `time' \(s\), the reflection is exponentially increased or diminished.  For example, simply translating (\ref{initial-profile}) by an imaginary distance \(u_{0}(x)\to u_{0}(x+{\rm i}x_{0})\) will change the reflection coefficient according to (\ref{rtrans}): \(r_{s}=r_{0}\exp(2k_{x}x_{0})\).  This alters the form of the profile around \(x=0\), continuously diminishing or amplifying the reflection until \(x_{0}=\pm\pi\sigma/2\) when one of the poles of the hyperbolic tangent is encountered, causing a jump in the reflectivity~\cite{horsley2016}.  For this particular profile we can therefore diminish the reflection by at most a factor of \(\exp(-2 \pi k_{x}\sigma)\) through translation, which requires gain in the material parameters.  Despite this practical limitation, this method of translating the profile by an imaginary distance to modify the scattering properties is promising, and may be similarly applied to two and three dimensional inhomogeneous profiles.
\par
Now consider evolution of (\ref{initial-profile}) by the third order KdV equation (\ref{kdvequation}), for simplicity over a short period of imaginary `time' \(s={\rm i}\tau\).  This results in a complex profile with an imaginary part in addition to the real part (\ref{initial-profile}),
\begin{multline}
	u_{s}(x)=\frac{U_0}{2}\left[1+\tanh(x/\sigma)\right]\\
	+\frac{{\rm i}a U_0\tau}{\sigma}{\rm sech}^{2}(x/\sigma)\left\{\left[\frac{3k_{0}^{2}U_0}{2}-\frac{1}{\sigma^{2}}\right]+3\tanh(x/\sigma)\left[\frac{1}{\sigma^{2}}{\rm tanh}(x/\sigma)+\frac{k_{0}^{2}U_0}{2}\right]\right\}\label{evolved-profile}
\end{multline}
and has a reflection coefficient which is diminished by a factor of \(\exp(-8 a k_{x}^{3}\tau)\) (see figure~\ref{reflection-fig}), although we must remember that \(\tau\) is restricted to take values such that the imaginary part of (\ref{evolved-profile}) is much smaller than the real part~\footnote{Note that the conservation law \(\partial_{s}\int u_{s} dx=0\) is not obeyed by (\ref{evolved-profile}) this is because \(u_{s}\) does not tend to the same constant at \(x=\pm\infty\).}.  In the limit of a sharp interface, \(1/\sigma^{2}\ll 3k_{0}^{2}U_{0}/2\) the imaginary part of (\ref{evolved-profile}) takes negative values, requiring the medium to have gain.  Meanwhile in the opposite limit of a slowly graded interface \(1/\sigma^{2}\gg 3k_{0}^{2}U_{0}/2\), the imaginary part is always positive.  The crossover between these two cases occurs when \(\sigma=2k_{0}^{-1}(1-\sqrt{2/3})^{1/2}/\sqrt{U_0}\sim0.13\lambda/\sqrt{U_{0}}\), which is the smallest scale of transition where gain is not required in the material.  This distributed loss must be implemented quite precisely in order to have the desired effect.
%
%: Summary
%
\section{Summary}
\par
There is a fascinating link between the Korteweg--de Vries (KdV) equation of fluid dynamics and the time independent Schr\"odinger equation: if the potential in the Schr\"odinger equation is `evolved' subject to the KdV equation then the reflectivity as a function of energy remains unchanged for every snapshot in `time'.  In the above work we have investigated the equivalent of this relationship for an electromagnetic wave propagating through an inhomogeneous slab of dielectric material, where the invariance of the reflectivity as a function of energy translates into invariance as a function of angle, for a fixed frequency.  
\par
The relationship between the KdV equation and the reflectivity of an inhomogeneous slab was derived using an alternative method to the usual operator one, where we considered the radiation due to the current induced in the material when the permittivity is changed by a small amount.  Through demanding that this additional radiation change the field at \(x\) in a way that depends only the field infinitesimally close to \(x\) we derived the first two equations of the KdV hierarchy, as well as the Kramers--Kronig media recently considered in~\cite{horsley2015,longhi2015,longhi2016,horsley2016}.  It may be that this provides an different physical picture of the link between the KdV hierarchy and wave propagation through inhomogeneous media.
\par
There is much more freedom to manipulate the reflectivity of a medium if the permittivity is allowed to take complex values.  Using an exact solution of the KdV equation (a limiting case of which is the P\"oschl--Teller potential) we found a very large class of both bounded and unbounded complex permittivity profiles that do not reflect from either side, for any angle of incidence, verifying this numerically.  In the special case of the periodic `cnoidal wave' solution to the KdV equation we found that the non--reflection property translates into there being a real value of the Bloch vector for all angles of propagation.  Finally we applied the evolution equations discussed in the first part of the paper to reduce the reflection of a wave from an interface, where the permittivity is rapidly increased from one constant value to another.  In this case the KdV equation can be used to find a distribution of the dissipative response that must be added to the interface in order that the reflectivity is decreased for all angles of incidence (although this reduction is much less for the angles far away from normal incidence).
\par
The overall purpose of this paper has been twofold: firstly to show that knowledge of the relationship between the KdV hierarchy and the Helmholtz equation can be a useful addition to existing design tools such as transformation optics, and provides another method for controlling scattering from complex inhomogeneous media.  The second purpose is to propose that metamaterials may be suitable for experimentally investigating the equivalence of scattering for media that have been `evolved' according to the KdV equation, a beautiful connection that has not yet been practically demonstrated.
%
%: Acknowledgements
%
\acknowledgements
The author wishes to thank T. G. Philbin, C. G. King, J. R. Sambles and A. P. Hibbins for illuminating conversations.
%
%: References
%

\end{document}